\documentstyle{mn} 

\newif\ifAMStwofonts

\input{epsf}   %% for postscript files

\newcommand{\acir}{$\alpha$~Cir}
\newcommand{\ha}{H$\alpha$}
\newcommand{\ms}{\,ms$^{-1}$}
\newcommand{\kms}{\,kms$^{-1}$}
\newcommand{\mh}{\,$\mu$Hz}
\newcommand{\mm}{\,mmag}

\title[Spectroscopy of \acir\ -- II.\ The \ha\ line]
{Spectroscopy of the roAp star $\balpha$~Cir -- II. \\
The bisector and equivalent-width of the H$\balpha$ line}
\author[I. K. Baldry et al.]
       {I. K. Baldry$^1$, M. Viskum$^2$,  T. R. Bedding$^1$,
                          H. Kjeldsen$^{2,3}$, S. Frandsen$^2$ \\
        $^1$Chatterton Astronomy Department, School of Physics, 
            University of Sydney, NSW 2006, Australia\\
        $^2$Institute of Physics and Astronomy, 
            University of Aarhus, DK-8000 Aarhus C, Denmark\\
        $^3$Theoretical Astrophysics Center, 
            University of Aarhus, DK-8000 Aarhus C, Denmark}
\date{Accepted 1998 September. 
      Received 1998 September; 
      in original form 1998 May.} 

\pagerange{\pageref{firstpage}--\pageref{lastpage}}

\pubyear{1997}

\begin{document}

\maketitle

\label{firstpage}

\begin{abstract}

We present bisector measurements of the \ha\ line 
of the rapidly oscillating Ap (roAp) star, $\alpha$~Cir, 
obtained from dual-site observations with medium-dispersion spectrographs. 
The velocity amplitude and phase of the principal pulsation mode 
vary significantly, depending on the height in the \ha\ line, 
including a phase reversal between the core and the wings of the line. 
This supports the theory, suggested in Paper~I, of a radial 
pulsational node in the atmosphere of the star. 
Blending with metal lines partially affects the \ha\ bisector results 
but probably not enough to explain the phase reversal. 

We have also detected changes in the equivalent-width of the line 
during the pulsation, and measured the oscillatory signal as a 
function of wavelength across the \ha\ region. 

\end{abstract}


\section{Introduction}

Rapidly oscillating Ap (roAp) stars are a sub-class of chemically 
peculiar magnetic (Ap or CP2) stars. 
They pulsate with photometric amplitudes below 8\mm\ 
and periods in the range 5--15 minutes. 
\acir\ (HR 5463, HD 128898) 
is the brightest of the known roAp stars ($V = 3.19$). 
Previous observations of this star in photometry 
have shown that it has one dominant pulsation mode 
with a period of 6.825\,min ($f = 2442$\mh ; 
Kurtz et al.\ 1994, hereafter KSMT). 
Medupe \& Kurtz (1998) observed that the amplitude decreased 
with increasing wavelength, from $2.71 \pm 0.18$\mm\  in 
Johnson~$U$ (3670\AA) to $0.41 \pm 0.13$\mm\ in $I$ (7970\AA). 
They proposed that the rapid decline of photometric amplitude with wavelength 
could be explained by a decrease in the temperature amplitude of the pulsation 
with atmospheric height.

Baldry et al.\ (1998a, hereafter Paper~I) showed that the velocity amplitude 
and phase of the principal pulsation mode in \acir\  varied significantly 
from line to line. 
However, it was difficult to interpret the data because of blending effects. 
In particular, 
there was evidence for a velocity node in the atmosphere of \acir\ 
but it was uncertain whether the node was horizontal or radial.
In this paper we look at the \ha\  line in more detail 
using the same set of observations, taken during two weeks in 1996 May.
These observations comprise 6366 intermediate-resolution spectra taken using 
the 74~inch (1.88-m) Telescope at Mt.~Stromlo, Australia 
and the Danish 1.54-m Telescope at La~Silla, Chile 
(see Section~2 from Paper~I for further details). 

We first examined how the \ha\  profile changed 
during the principal pulsation cycle, 
in terms of the bisector at different heights in the line 
(preliminary results were presented by Baldry et al.\ 1998b). 
In this way, the effect of the velocity on the profile could be analysed. 
To quantify the temperature effect, 
we measured the equivalent-width (EW) amplitude of the principal mode 
in filters of varying width and 
measured pixel-by-pixel intensity changes 
across the \ha\  region of the spectrum. 
Finally, we defined an observable quantity (related to the EW of \ha) 
which had a high signal-to-noise ratio for the principal pulsation. 
Using this observable, some of the weaker modes in \acir\  were detected.

\subsection{General Properties of \acir}
\label{sec:gen-prop}

Before we discuss the pulsation of \acir , 
we will review the general properties in the light of 
a recent spectral analysis (Kupka et al.\ 1996) 
and the Hipparcos parallax measurement (ESA 1997). 
The distance of $16.4 \pm 0.2$\,pc (parallax of $61.0 \pm 0.6$\,mas), 
combined with a bolometric correction of $-0.12 \pm 0.02$ 
($M_{\rm bol,\sun} = 4.64$, Schmidt-Kaler 1982), 
gives $M_{\rm bol} = 2.00 \pm 0.04$ ($L = 11.4 \pm 0.4 L_{\sun}$). 
{}From this luminosity and 
the temperature of $ T_{\rm eff} = 7900 \pm 200 $\,K 
(derived by Kupka et al.), 
we obtain $R = 1.81 \pm 0.11 R_{\sun}$ 
(angular diameter of $1.03 \pm 0.07$\,mas). 
Combining this radius with $\log g = 4.2 \pm 0.15$ (Kupka et al.) 
gives a mass of $M = 1.9 \pm 0.6 M_{\sun}$. 
The rotation period of 4.48 days (derived by KSMT) 
and the radius means that $v_{\rm rot} = 20.4 \pm 1.2$\kms . 
Using $ v \sin i = 13 \pm 1 $\kms (Kupka et al.), 
the inclination of the rotation axis 
to the line of sight is then $i = 40\degr \pm 5\degr$.
This is an improvement on the estimate given by KSMT due 
to more recent results. 

\subsection{The Oblique Pulsator Model}
\label{sec:obl-puls}

In some roAp stars, including \acir, 
the amplitude of the oscillations has been observed to 
be modulated by the rotation period of the star. 
In particular, the amplitude is at a 
maximum when the observed magnetic field strength is also at a maximum. 
This led to the oblique pulsator model (Kurtz 1982), 
in which an roAp star pulsates 
non-radially with its pulsation axis aligned with the magnetic axis. The 
amplitude modulation comes from the inclination of the pulsation axis with 
respect to the rotation axis. Therefore different aspects of a non-radial 
mode are viewed as the star rotates. The oblique pulsator model predicts that,
in the Fourier domain, a frequency of mode $\ell$ is split into 2$\ell$\,+1 
frequencies. 
The frequency splitting is exactly equal to the rotation frequency, 
and the relative amplitudes are determined by both 
the inclination of the rotation axis to our line of sight ($i$) and 
the angle between the rotation axis and the pulsation axis ($\beta$). 
Most of the review papers mentioned in the Paper~I 
discuss the oblique pulsator model, a refinement of which 
is given by Shibahashi \& Takata (1993). 

The principal pulsation mode in \acir\ is believed to be a pure oblique 
dipole mode ($\ell = 1$). 
KSMT observed a 21 percent full-range variation of the amplitude 
during the rotation period. 
This variation implies that 
$\tan i \tan \beta = 0.21 \pm 0.01$ (see Section~4.1 from KSMT). 
Using our new estimate of $i$ (see Section~\ref{sec:gen-prop}), 
we get $\beta = 14\degr \pm 3\degr$. 
This means that 
the inclination angle of the pulsation axis to the line of sight ($\alpha$) 
varies between 26\degr\  and 54\degr\  during the rotation cycle.

\section{General Data Reductions}

Extraction of spectra and continuum fitting were done 
using IRAF procedures (see Section~3.1 from Paper~I). 
We made a number of types of measurements on the reduced spectra, 
including bisector line-shift measurements which are described in 
Section~\ref{sec:reduc-bis-wid}. 
The time series analysis of these is explained in Section~\ref{sec:time-ser} 
and is also applicable to other measurements, 
which are described later in the paper. 
Since the continuum fit is more critical than for Paper~I, 
we first describe this procedure in more detail. 

\subsection{Continuum Fitting}
\label{sec:cont-fit}

The IRAF procedure {\tt continuum}
was applied to each spectrum in our data, using the following parameters: 
{sample=@list, naverage=1, function=legendre, 
order=3/4 `meaning 2nd/3rd order polynomial', low\_rej=2, high\_rej=4, 
niterate=20/25, grow=0}. 

Initially, a sample of about 1400 points from each spectrum was used 
to make a least-squares polynomial fit --- 3rd order for the Stromlo data 
and 2nd order for the La Silla data. A higher order fit was used for the 
Stromlo data because the continuum shape was less stable. 
Next, points below 2$\sigma$ and above 4$\sigma$ from the fit were excluded 
from the next fit. 
The lower cutoff was used to exclude some absorption lines and 
the upper cutoff only excluded cosmic ray events. 
The fitting and excluding routine was repeated 20 to 25 times or in most cases 
until no more points were excluded. 
In the final fit, about 1000 points were included, 
with none being within 50\AA\ of the core of the \ha\  line. 
Not all absorption lines were excluded from the final fit 
since there is very little real continuum at the resolution 
of our data ($\sim$1.5\AA). 
The real continuum level in each fitted spectrum was around 1.005--1.010, 
depending on the wavelength. 
Therefore, we have scaled the spectra so that 
the continuum level is $\sim$1.00 in the \ha\  region of the spectrum.
The level varied slightly from spectrum to spectrum, 
with a standard deviation of about 0.001. 

\subsection{Bisector Measurements of \ha}
\label{sec:reduc-bis-wid}

\begin{figure}
\epsfxsize=9.5cm
\centerline{\epsfbox{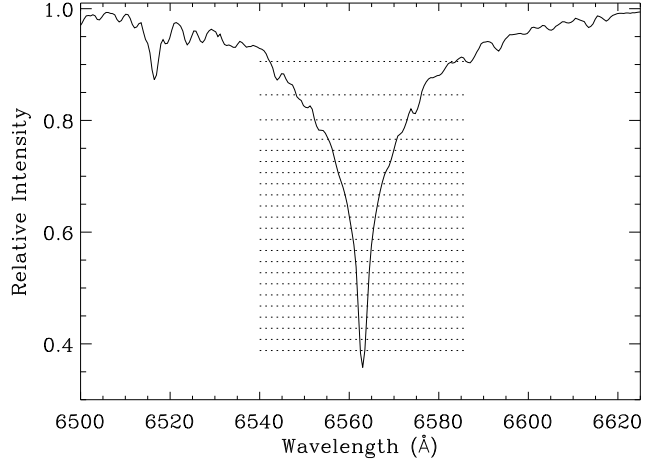}}
\caption{The \ha\  line in \acir . The dotted lines divide 
the 22 contiguous sections.}
\label{fig:sp-ha}
\end{figure}

\begin{table*}
\caption{Amplitudes and phases of the principal pulsation mode 
for the bisector velocity at different heights in the \ha\ line. 
The line was divided into 22 non-overlapping and two extra horizontal sections.
For the results in this table, 
the Stromlo and the La Silla data sets were combined.} 
\label{tab:sections}
\begin{tabular}{rrrrrrrrr} \hline
section & low    & high   & mean             &
velocity  & velocity  & S/N & velocity  & phase   \\
  no.   & cutoff\rlap{$^a$} & cutoff\rlap{$^a$} & width\rlap{$^b$} &
ampl.\rlap{$^c$} & noise\rlap{$^d$} & & phase\rlap{$^e$} & error\rlap{$^f$} \\
        &        &        & (\AA)     &
(\ms)     & (\ms)     &     & (\degr)   & (\degr)   \\
 \\
 0&  0.39& 0.41&  1.1& 264& 19& 14.1& 333&  4\\ 
 1&  0.41& 0.43&  1.4& 256& 15& 17.4& 328&  3\\ 
 2&  0.43& 0.45&  1.7& 279& 15& 18.8& 327&  3\\ 
 3&  0.45& 0.47&  1.9& 277& 15& 18.7& 327&  3\\ 
 4&  0.47& 0.49&  2.2& 257& 15& 16.9& 328&  3\\ 
 5&  0.49& 0.51&  2.5& 227& 15& 14.7& 330&  4\\ 
 6&  0.51& 0.53&  2.7& 193& 16& 11.9& 333&  5\\ 
 7&  0.53& 0.55&  3.0& 176& 18&  9.6& 331&  6\\ 
 8&  0.55& 0.57&  3.5& 125& 18&  6.9& 331&  8\\ 
 9&  0.57& 0.59&  3.9&  43& 18&  2.4& 303& 24\\ 
10&  0.59& 0.61&  4.6&  54& 20&  2.8& 255& 21\\ 
11&  0.61& 0.63&  5.4&  55& 21&  2.6& 283& 22\\ 
12&  0.63& 0.65&  6.2&  43& 23&  1.9& 334& 32\\ 
13&  0.65& 0.67&  7.1&  26& 25&  1.0& 304& 73\\ 
14&  0.67& 0.69&  8.3&  64& 29&  2.2& 205& 27\\ 
15&  0.69& 0.71&  9.7&  25& 33&  0.8& ---&---\\
16&  0.71& 0.73& 11.6& 254& 37&  6.8& 195&  8\\ 
17&  0.73& 0.75& 13.0& 340& 34& 10.1& 192&  6\\ 
18&  0.75& 0.77& 14.4& 310& 36&  8.5& 191&  7\\ 
19&  0.77& 0.80& 19.0& 181& 45&  4.0& 215& 14\\ 
20&  0.80& 0.85& 23.9& 514& 44& 11.7&  45&  5\\ 
21&  0.85& 0.91& 32.2& 208& 56&  3.7&  18& 16\\ 
 \\
22&  0.70& 0.72& 10.5&  86& 36&  2.4& 200& 25\\ 
23&  0.76& 0.78& 15.3& 221& 45&  4.9& 195& 12\\ 
 \hline
\end{tabular}
\begin{flushleft}
$^{a}$Boundary of the section in relative intensity.
\newline
$^{b}$Approximate mean width of the line at the height of the section. 
\newline
$^{c}$Amplitude measured at 2442.03\mh .
\newline
$^{d}$rms-noise estimated from amplitude spectrum using the regions 
1100--2300\mh\ and 2600--4400\mh .
\newline
$^{e}$Phase measured at 2442.03\mh , 
with respect to a reference-point ($t_{0}$) at JD 2450215.07527, 
with the convention that a phase of 0\degr\ represents 
maximum of the observed variable. 
\newline
$^{f}$Error in the phase is taken to be arcsin\,(rms-noise/amplitude). 
\newline
\end{flushleft}
\end{table*}

The \ha\  line in each spectrum was divided 
into 22 contiguous horizontal sections 
(see Fig.~\ref{fig:sp-ha} and Cols.~1--4 of Table~\ref{tab:sections}) 
and two extra sections for checking. 
For each section, 
the average wavelength of each side of the absorption line was measured, 
and a bisector line-shift (average position of the two sides) 
was calculated. 
We have used a telluric O$_{2}$ band as a Doppler shift reference 
(see Section~3.3 from Paper~I for details). 

An alternative method was also tried in which a least-squares fit was made 
to the position of each side of the line. 
A template spectrum was used to define the shape of the line 
and was shifted from side to side until a best fit was obtained. 
This method produced similar results and noise levels to the 
method of calculating the average wavelength, 
but the computational time was longer. 

\subsection{Time Series Analysis}
\label{sec:time-ser}

For each section of the \ha\  line, the above analysis produced 
a time series of 6366 bisector line-shift measurements 
(4900 from Stromlo, 1466 from La Silla). 
Each time series was high-pass filtered 
and then cleaned for bad data points by removing any points lying outside 
$\pm 6.5$ times the median deviation. 
Typically, about 100 data points were removed. 
Next, a weighted least-squares sine-wave fitting routine was applied 
(using heliocentric time) to produce amplitude spectra. 

For the analysis of the principal pulsation mode in \acir , 
we have measured the amplitude and phase of each time series at 
2442.03\mh\  and estimated the rms-noise level by averaging over 
surrounding frequencies, 1100--2300\mh\  and 2600--4400\mh .
The phases are measured at a temporal phase reference point ($t_{0}$) 
with the convention that a phase of 0\degr\ represents maximum 
of the observable. $t_{0}$ was chosen to coincide with maximum light 
using data supplied by Don Kurtz (private communication); 
see Paper~I for details. 
For the phase error, 
we have used simple complex arithmetic to find the maximum change in phase 
that an rms-noise vector could induce, i.e., arcsin\,(rms-noise/amplitude). 

\section{Bisector Velocities}

\subsection{Results}
\label{sec:results-bis}

\begin{figure}
\epsfxsize=9.5cm
\centerline{\epsfbox{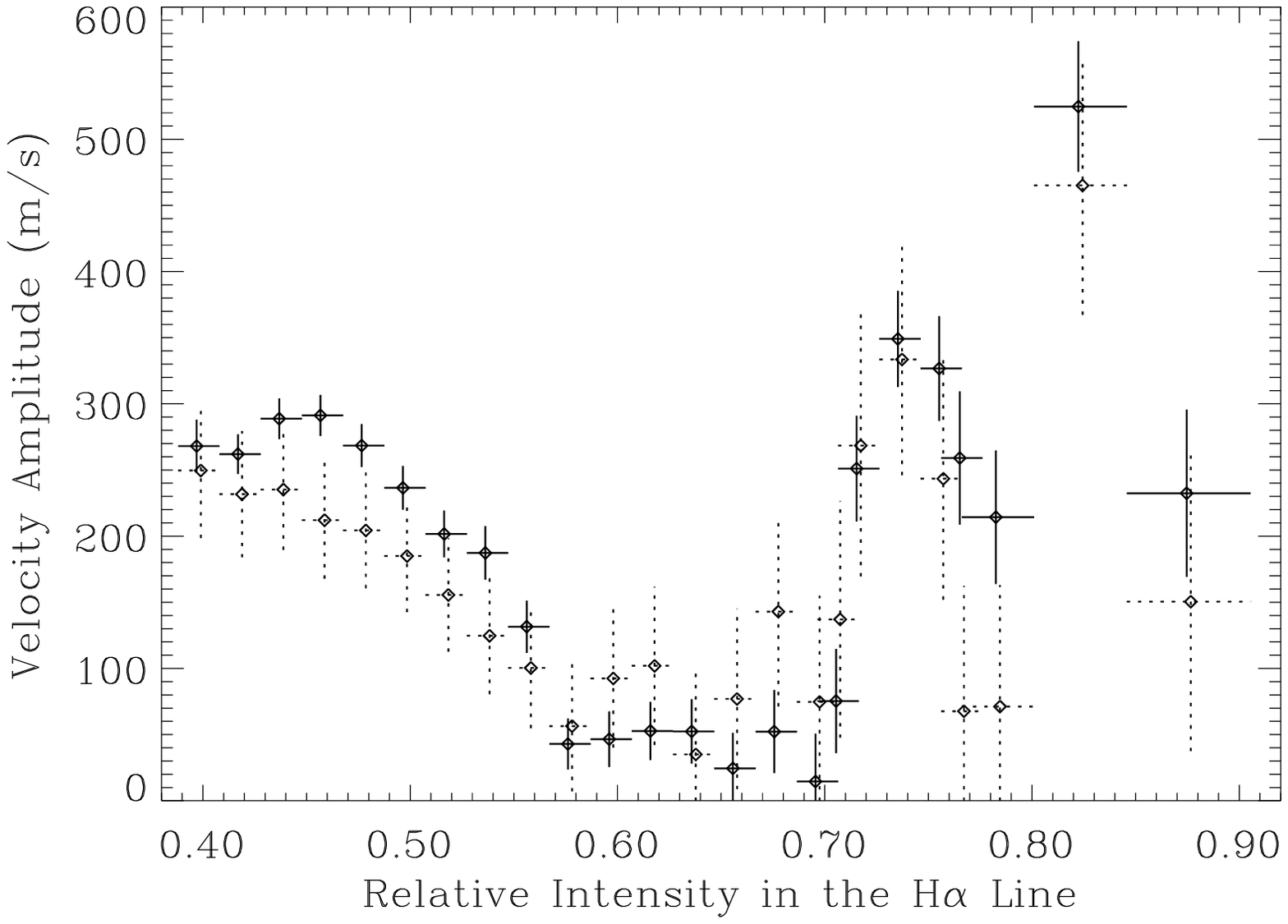}}
\epsfxsize=9.5cm
\centerline{\epsfbox{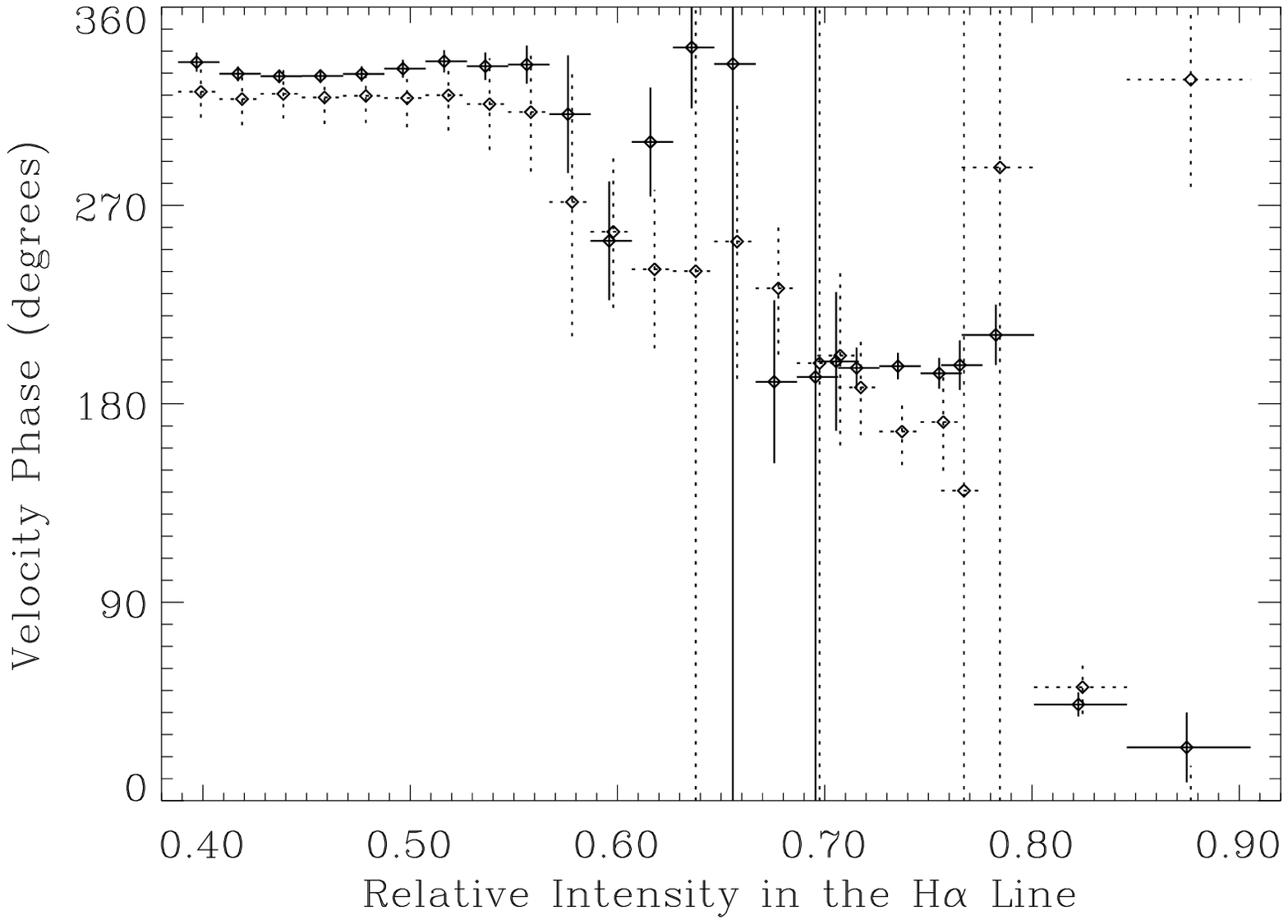}}
\caption{Amplitudes and phases of the principal pulsation mode for 
the bisector velocity at different heights in the \ha\  line. 
Points with solid lines represent the Stromlo data and 
points with dotted lines represent the La Silla data. 
For each measurement, the vertical line is an error-bar while 
the horizontal line shows 
the extent of the section in the \ha\  line.}
\label{fig:bis-vel}
\end{figure}

\begin{figure}
\epsfxsize=9.5cm
\centerline{\epsfbox{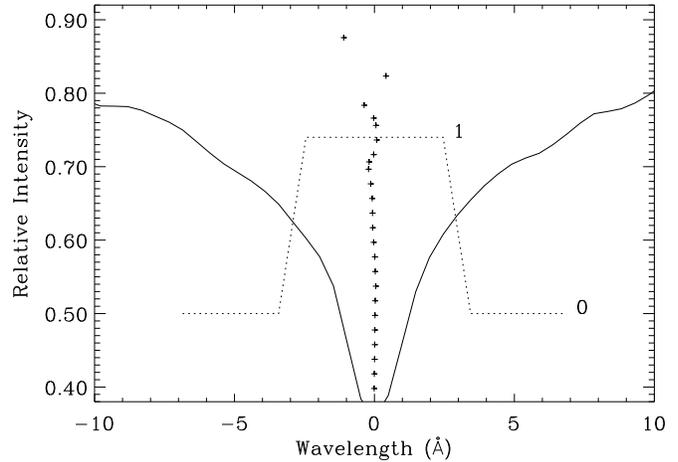}}
\caption{The \ha\  line in \acir . The crosses show the mean position 
of the bisector. The wavelength is measured with respect to 
the centre of the line. 
The dotted line shows the shape of trapezium filter no.~2, 
which is used for the equivalent-width measurements. 
It is plotted using a different vertical scale.}
\label{fig:bis-ha}
\end{figure}

We have assumed that the bisector line-shift measurements represent 
velocities in the star. 
The velocity amplitudes and phases of the principal mode 
at different heights in the \ha\  line are shown 
in Fig.~\ref{fig:bis-vel} and Table~\ref{tab:sections}. 
The results describe the oscillations in the bisector 
about the mean position at each height 
(see Fig.~\ref{fig:bis-ha} for the bisector shape). 
{}From 0.4 to 0.8 in the line, 
the amplitude decreases from 300\ms\  to zero and then increases again, 
with a change in phase of $-$140\degr . 
Note that there is good agreement between the Stromlo and La~Silla data sets, 
which were analysed separately for Fig.~\ref{fig:bis-vel}. 
This gives us confidence that the observed variations are intrinsic to the 
star. 

In Paper~I, the \ha\  velocity amplitude and phase were measured to be 
$\sim$170\ms\  and $\sim$330\degr\  using a cross-correlation method. 
This is compatible with our bisector velocity results since the 
cross-correlation measurement would be dominated by the steepest part 
of the \ha\  profile, near the core. 

\begin{figure}
\epsfxsize=9.5cm
\centerline{\epsfbox{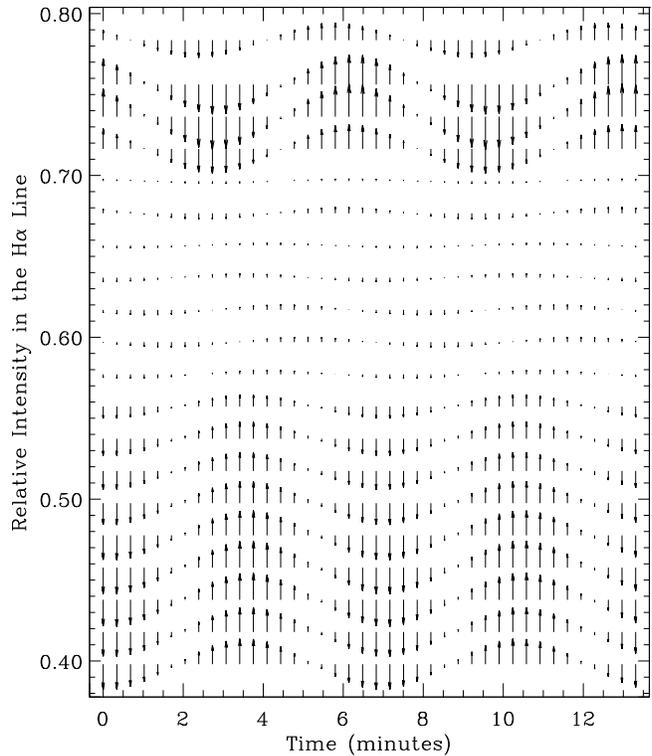}}
\caption{Velocity field diagram. The arrows represent bisector velocity 
vectors, relative to the mean velocity of the star, 
at different times throughout two 
pulsation cycles of the principal mode ($P=6.825$\,min). 
The velocity vectors are calculated from 
the amplitudes and phases of sections~0--19 given in Table~\ref{tab:sections}.}
\label{fig:vel-field}
\end{figure}

To further illustrate the behaviour of the \ha\  bisector, 
in Fig.~\ref{fig:vel-field} 
we show the bisector velocity as a function of time and height. 
The horizontal axis covers two cycles of 
the principal pulsation mode ($P=6.825$\,min). 
Only sections 0--19 are shown (see Table~\ref{tab:sections}), 
since above a height of 0.8 the measurements are significantly affected 
by line blends (see Section~\ref{sec:blend}). 
The figure clearly illustrates that the higher sections are pulsating nearly 
in anti-phase with the lower sections, and that the velocities of the middle 
sections are close to zero. 
Since the bisector velocity reflects the velocity at different heights 
in the atmosphere, 
these results support the hypothesis, suggested in Paper~I, 
of a radial node in the atmosphere of \acir . 

\subsection{Blending Considerations}
\label{sec:blend}

We attribute the velocity node (at $\sim$0.65) and 
the phase jump between 0.4 and 0.8 to \ha\  line-formation effects, 
although there is some uncertainty as to how much blending affects the 
results between heights 0.7 and 0.8. 
Above a height of 0.8, the results are significantly affected by line blending.
In order to identify any metal lines in the wings of \ha , 
we used a synthetic spectrum calculated by 
Friedrich Kupka (private communication) 
and high-resolution spectra of \acir\ taken in 1997 March 
(coud\'{e} echelle, 74~inch Telescope, Mt.~Stromlo). From this, 
we determined that most of the lines blended with the \ha\ profile 
from 6540\AA\ to 6585\AA\ are telluric, 
including all those below a height of 0.7 in the \ha\ line.

At height 0.82, the anomalously large velocity amplitude is probably caused by 
the metal absorption line Sr~I 6550.2\AA , 
with the possibility that Fe~I 6575.0\AA\ also contributes. 
Since metal lines can have amplitudes as large as 1000\ms\  (Paper~I), 
this can explain the large bisector velocity seen at this height. 
At height 0.88, the amplitude is similar to that between 
0.72 and 0.79 but the phase is anomalous. 
This measurement is affected by various metal and telluric lines, 
in particular a metal line blend including 
Mg~II 6546.0\AA\ and Fe~I 6546.2\AA .

Below a height of 0.8, we expect any absorption features to have 
less impact on the bisector velocity due to the increasing steepness 
of the \ha\  profile.
However, there is a metal line Fe~I 6569.2\AA\ in addition 
to a few telluric lines that affect the \ha\ profile between 0.7 and 0.8. 
This Fe~I feature affects the bisector measurements between 0.71 and 0.76. 
Therefore, it probably causes the jump in amplitude near height 0.72 
shown in Fig.~\ref{fig:bis-vel}.
It cannot explain why all the bisector velocities 
from heights 0.67--0.80 have phases around 190\degr --200\degr , 
but it does lower the significance of the phase jump between the 
core and this part of the wings. 
We speculate that without this Fe I feature, 
the \ha\ bisector velocity amplitude would increase steadily from 0.70 to 0.77.

In summary, we believe features in Fig.~\ref{fig:bis-vel} below height 0.7 
reflect genuine effects in the \ha\  profile, 
while those above 0.8 are dominated by metal and telluric lines. 
Between 0.7 and 0.8, a blend probably exaggerates the phase reversal by 
causing an increase in the measured amplitude around height 0.73. 

\subsection{Simulations}
\label{sec:simul}

Two sources of noise for the velocity measurements, 
especially at high levels in the \ha\  line, 
are errors in the continuum fit and changes in the total 
equivalent-width (EW) of the line 
(which are expected due to temperature changes during the pulsation cycle). 
These sources will affect the bisector velocity if 
the line is asymmetrical at that height. 
To investigate, 
we first simulated fluctuations in the EW of the line and
measured the resulting pseudo-velocity amplitude, 
and then simulated fluctuations in the continuum level.
We concluded that, while there could be systematic changes in the 
velocity amplitude greater than 150\ms\  above a height of 0.65, 
the phase of 
the systematic change would be opposite between height 0.72 and 0.76. 
This means that this effect cannot explain 
the different phase of heights 0.72--0.79 compared to heights 0.40--0.56. 
Perhaps, this systematic effect could explain the large difference 
in velocity amplitude between height 0.70 and 0.72. 

We also considered the effect of a systematic change 
in the slope of the continuum across the \ha\  line. 
This could arise because the continuum fit is not entirely independent 
of the metal lines and, 
since there are significantly more metal lines on the blueward side of \ha , 
the slope of the continuum may vary during the pulsation cycle. 
We have simulated the effect on the velocity amplitude of 
a continuum variation having a slope change of 0.001 / 100\AA . 
We found that such a large slope change could produce a systematic change in 
the velocity amplitude greater than 300\ms\  above a height of 0.75 and thus 
mimic a velocity phase reversal. 
However, the simulation fails to reproduce a velocity node at 0.65 and 
other features of Fig.~\ref{fig:bis-vel}. 
Furthermore, from calculations of the intensity variations in various regions 
of the spectrum, 
we do not expect any systematic continuum slope changes to be larger 
than 0.0001 / 100\AA .

Our simulations show that continuum level, continuum slope and \ha\ EW changes 
cannot account for the observed features of Fig.~\ref{fig:bis-vel}. 
Therefore, we conclude that the node and phase reversal 
are caused by the velocity field of the star. 
Note that random variations in the continuum level and slope may be larger 
than tested for, 
but this will only effect the noise level in the amplitude spectra.

\subsection{Discussion}
\label{sec:discuss-bis}

Hatzes (1996) has simulated line-bisector variations for non-radial 
pulsations in slowly rotating stars. His simulations show that a 
bisector velocity phase reversal could occur in modes with $\ell = m \ge 3$ 
(see Fig.~3 from Hatzes 1996). 
In fact, the \ha\  line-bisector variations in \acir\ 
could closely be described by his simulations with $\ell = m = 3$~or~4.
However, he does not take into account any changes in pulsational 
amplitude with depth and the simulations were done for much narrower 
metal lines. 
Given that there is strong evidence for the $\ell = 1$ oblique 
pulsation model (KSMT), we suggest that a change in pulsational 
amplitude with depth is a more likely explanation for the bisector variations. 

It is interesting to note that similar behaviour to that seen in \acir\ 
has been seen in the Sun. 
Deubner, Waldschik \& Steffens (1996) 
have observed a phase discontinuity in 
the solar 3-min oscillations using spectroscopy. 
In particular, they measured phase differences 
between the velocity of the core of the NaD$_{2}$ line and 
various positions in the wing (called V$-$V phase spectra). 
They discovered a 180\degr\  phase jump in the V$-$V spectra near 
a frequency of 7000\mh . 
They also observed a phase discontinuity in the 
V$-$I (Line Intensity) spectra at a similar frequency. 

The 3-min oscillations are thought to be formed 
in an atmospheric (or chromospheric) cavity. 
Deubner et al.\ (1996) suggested a model to explain the phase discontinuities, 
involving running acoustic waves and atmospheric oscillation modes, 
one of which must have a velocity node in the observed range of heights 
in the atmosphere. 

In the solar model, the eigenfunction of a p-mode with $\ell = 1$ and 
$n = 25$ has a radial node separation in the outer part of the Sun of 
$\sim$0.3 percent of the radius (Christensen-Dalsgaard 1998). 
In \acir , if we assume the oblique dipole pulsation model ($\ell = 1$) for 
the principal mode, with a large frequency separation 
$\Delta \nu_{0} = 50$\mh\ (KSMT), then the overtone value of the 
principal mode is $n = 48$, 
using the asymptotic theory of asteroseismology 
(e.g.\  Brown \& Gilliland 1994). 
The radial node separation in \acir\ is then expected to be $\sim$0.15 
percent of the radius of the star. 
This is equivalent to a distance of about 1900\,km, 
assuming the radius of \acir\ is about twice the radius 
of the Sun (see Section~\ref{sec:gen-prop}).
This does not take in to account 
the difference in density as a function of radius between \acir\  and the Sun.
Using model atmospheres (Friedrich Kupka, private communication) 
to estimate the sound speed ($v$) just above the photosphere, 
we obtain a radial node separation of about 1500\,km ($v/2f$). 
This is in good agreement with the estimate obtained by scaling from a 
solar model. 

In the model of \acir\  used by Medupe \& Kurtz (1998), 
the geometric height between the formation of the $I$ band and the $B$ band 
is 250\,km. We speculate that the extent of the line forming region 
is about 1000\,km and that we are seeing one velocity node in the atmosphere. 
Recently Gautschy, Saio \& Harzenmoser (1998) described pulsation models 
for roAp stars which suggest that radial nodes can be 
expected in the atmospheres of these stars. 

\section{Equivalent-width Measurements of \ha}

Kjeldsen et al.\ (1995) developed a new technique 
for detecting stellar oscillations 
through their effect on the equivalent-width (EW) of Balmer lines. 
We expect to find changes in the \ha\ EW in \acir\ due to temperature changes 
in the star. 

\subsection{Reductions}
\label{sec:ew-reduc}

We measured the EW changes of the \ha\ line directly 
by looking at intensity changes in regions of varying width across \ha .
First of all, the spectra were linearly re-binned by a factor of 40 and 
shifted so that the centre of the \ha\  line was in the same wavelength 
position for each spectrum. 
This was to reduce the noise caused by instrumental shifts of the spectra. 
The value used to define the centre of the \ha\  line in each spectrum 
was taken from the cross-correlation measurements used for Paper~I.
Next, the mean intensities in 11 trapezium-shaped filters 
(Cols.~1--2 of Table~\ref{tab:ew-meas}) were measured in each spectrum. 
Each filter was centred on \ha\  and the sloping part of the trapezium 
was two pixels wide ($\sim 1$\AA) at each end 
(see Fig.~\ref{fig:bis-ha} for an example). 

\begin{table*}
\caption{Equivalent-width (EW) amplitudes and phases of the principal mode 
for filters of different width across \ha . 
The measured relative intensity amplitudes ($\delta I$) 
are converted to fractional EW amplitudes ($\delta W / W$), 
using the formula given in Section~\ref{sec:ew-reduc}.
For the results in this table, 
the Stromlo and the La Silla data sets were combined.}
\label{tab:ew-meas}
\begin{tabular}{rrrrrrrrrr} \hline
filter & filter & mean & 
intensity & intensity & EW & EW & 
S/N & EW & phase \\
no. & width\rlap{$^a$} & intensity\rlap{$^b$} & 
ampl.\rlap{$^c$} & noise\rlap{$^d$} & ampl.\rlap{$^c$} & noise\rlap{$^d$} & 
    & phase\rlap{$^e$} & error\rlap{$^f$} \\
    & (\AA)  &    & 
(ppm) & (ppm) & (ppm) & (ppm) &
    & (\degr) & (\degr) \\ 
 & & & & & & & & & \\
  0 &   1.0 & 0.378 & 1085 & 72 & 1743 & 116 & 15.1 & 354 &  4 \\
  1 &   2.0 & 0.402 &  996 & 42 & 1667 &  69 & 24.0 & 354 &  2 \\
  2 &   5.9 & 0.511 &  692 & 20 & 1415 &  40 & 35.4 & 357 &  2 \\
  3 &  13.7 & 0.616 &  441 & 20 & 1148 &  51 & 22.6 &   1 &  3 \\
  4 &  17.6 & 0.650 &  390 & 20 & 1114 &  56 & 19.7 &   3 &  3 \\
  5 &  24.5 & 0.694 &  301 & 20 &  982 &  65 & 15.0 &   1 &  4 \\
  6 &  31.9 & 0.730 &  246 & 20 &  913 &  76 & 12.1 &   7 &  5 \\
  7 &  45.1 & 0.779 &  201 & 21 &  909 &  95 &  9.6 &   9 &  6 \\
  8 &  66.6 & 0.829 &  156 & 21 &  911 & 122 &  7.5 &  13 &  8 \\
  9 &  81.8 & 0.853 &  147 & 21 &  995 & 141 &  7.1 &  12 &  8 \\
 10 & 110.7 & 0.882 &  122 & 20 & 1039 & 172 &  6.0 &  11 & 10 \\ \hline
\end{tabular}
\begin{flushleft}
$^{a}$Full-width half-maximum of filter centred on \ha . 
\newline
$^{b}$Mean relative intensity in the filter averaged over all the spectra. 
The continuum level is $\sim$1.00. 
\newline
$^{c,d,e,f}$See Table~\ref{tab:sections}.
\newline
\end{flushleft}
\end{table*}

For each filter, we obtained a time series of intensity measurements 
which was analysed in the same way as in Section~\ref{sec:time-ser} 
to yield the amplitudes shown in Cols.~4--5 of Table~\ref{tab:ew-meas}.
{}From an intensity change ($\delta I$), 
we defined the fractional EW change as: 
\[
\frac{\delta W}{W} = \frac{- \delta I}{C - I}
\]
where $W$ is the equivalent-width, $C$ is the continuum level 
(approx.\ 1.00 in our reduced spectra) 
and $I$ is the mean intensity in the filter averaged over all the spectra 
(Col.~3 of Table~\ref{tab:ew-meas}). 
Using this formula, we converted intensity amplitude spectra to EW amplitude 
spectra. 
To test whether any bisector variations could affect these measurements, 
we simulated shifts of $\pm$600\ms\ in the template spectrum. 
From this, we determined that the effect of such a signal 
on the filter intensities would be less than the rms-noise level. 

\subsection{Results}
\label{sec:ew-res}

\begin{figure}
\epsfxsize=9.5cm
\centerline{\epsfbox{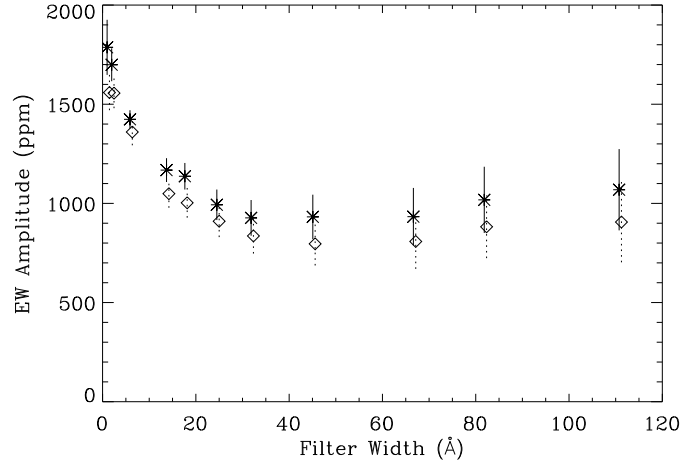}}
\caption{The EW amplitude in filters of different width 
centred on the \ha\  line. 
The stars with solid lines represent the Stromlo data 
and diamonds with dotted lines represent the La Silla data.}
\label{fig:ew-filter}
\end{figure}

\begin{figure}
\epsfxsize=9.5cm
\centerline{\epsfbox{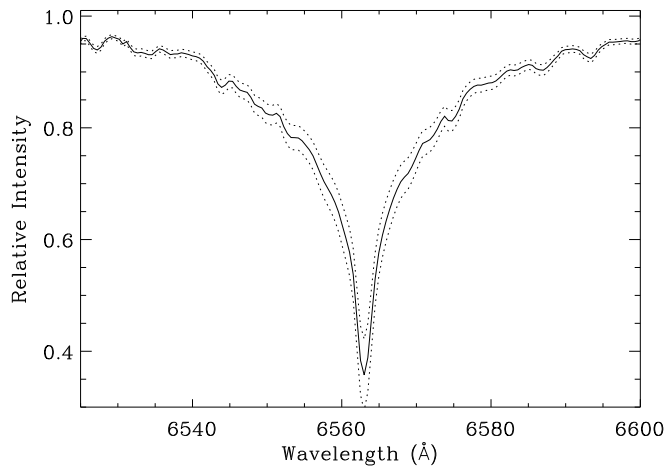}}
\caption{The \ha\  line with the dotted lines showing the profile with an 
increase and a decrease of 10 percent in EW. A variation of this type, 
but with much smaller amplitude (1000\,ppm), was used to generate the 
dashed line in Fig.~\ref{fig:cos-sin-wave}; 
and would produce a constant amplitude in Fig.~\ref{fig:ew-filter}.}
\label{fig:ha-sim}
\end{figure}

The results for the principal mode are shown 
in Cols.~6--10 of Table~\ref{tab:ew-meas} and in Fig.~\ref{fig:ew-filter}. 
The phases of all the measurements lie between $-$10\degr\  and 15\degr . 
Since our phase reference point ($t_{0}$) coincides with maximum light 
(see Section~4 from Paper~I), 
we conclude that the EW of the \ha\  line is pulsating 
in phase with the luminosity. 
This is expected, since maximum EW of the \ha\  line indicates maximum 
temperature in the stellar atmosphere. 
If the EW amplitude were the same in all filters then 
the change in intensity at each wavelength would be 
to be proportional to the depth of the absorption at that wavelength, 
i.e., $\delta I \propto C - I$. 
Such a profile variation is shown 
greatly exaggerated in Fig.~\ref{fig:ha-sim}. 
Our results in Fig.~\ref{fig:ew-filter} suggest that this is nearly the case, 
with an amplitude of 1000\,ppm, except that 
the core of the line is fluctuating in intensity by more than expected. 

We must remember that the measurements were made on continuum-fitted spectra, 
which means that the absolute flux\footnote
{
We use the term {\tt intensity} when referring to continuum-fitted spectra and 
the term {\tt flux} when considering the true flux from the star.
}
in the core might be constant while 
the wings and continuum are changing in flux. 
We do have information about the pulsation in continuum flux. 
Medupe \& Kurtz (1998) measured the photometric amplitude in Johnson $R$ 
(central wavelength 6380\AA) to be $0.54 \pm 0.15$\mm\  ($500 \pm 140$\,ppm). 
A continuum flux amplitude of 500\,ppm could only account 
for a 200\,ppm \ha\  core relative intensity amplitude 
in continuum fitted spectra. 
This value is 
significantly smaller than the measured value of $\sim$1000\,ppm (filter~1).
Therefore, we can say with some certainty that 
the continuum flux from the star is varying in anti-phase with 
the flux in the core of the line and that, 
at some point in the wings of \ha  , the flux is constant. 
To investigate this, 
we have calculated the intensity amplitude of the principal 
oscillation mode at each pixel in the spectrum, 
as we describe in Section~\ref{sec:pix-reduc}. 

\subsection{The Equivalent-width Amplitude}
\label{sec:ew-predict}

Using the photometric amplitude for \acir\ of $\sim$1.7\mm\ (Johnson $B$) 
measured by Don Kurtz (private communication) 
around the time of our observations, 
we can calculate the expected EW amplitude of the principal pulsation mode. 
The photometric amplitude converts to 
$( \delta L / L )_{\rm bol} = 1.5 \times 10^{-3}$ using the 
relation of Kjeldsen \& Bedding (1995) with $ T_{\rm eff} = 7900$\,K. 
In order to convert this to an EW amplitude, we use the scaling law from 
Bedding et al.\ (1996) which can be written as; 
\[
\frac{\delta W}{W} = 
\frac{1}{4}
\frac{\delta \ln W}{\delta \ln T_{\rm eff}} 
\left(\frac{\delta L}{L} \right)_{\rm bol}.
\]
Model atmospheres can be used to estimate the change in 
EW of the Balmer lines as a function of temperature. 
{}From Kurucz's models of the H$\alpha$ profile 
with log\,$g = 4.0$ (see Table~8A from Kurucz 1979), 
$\delta \ln W / \delta \ln T  \sim 2.8$ between 7500\,K and 8000\,K. 
Using this value, we predict an EW amplitude of 
$\delta W / W = 1.0 \times 10^{-3}$.

The measured amplitude of $1000 \pm 140$\,ppm 
(filter~9, see Table~\ref{tab:ew-meas}) 
is in excellent agreement with the prediction. 
However, to some extent the agreement is fortuitous because in reality, 
neither $( \delta L / L )_{\rm bol}$ nor 
$\delta \ln W / \delta \ln T$ are particularly well known. 
In the first case, since the photometric amplitude of \acir\  varies 
with wavelength by more than expected for a blackbody (Medupe \& Kurtz 1998), 
this will affect the bolometric luminosity amplitude. 
In the second case, the value depends significantly 
on the effective temperature and the accuracy of the model, 
giving a possible range in $\delta \ln W / \delta \ln T$ of 2--4. 
Additionally, 
there are uncertainties in the sensitivity of the EW amplitude to 
different modes (Bedding et al.\ 1996). 
Finally, note that the velocity amplitude is predicted to be 120\ms\ using the 
same scaling laws (Kjeldsen \& Bedding 1995), while the measured amplitudes 
vary between 0 and 1000\ms\ (Paper~I). 

\section{Pixel-by-Pixel Intensity Measurements}

Another way of examining changes in the \ha\ profile is to look at  
relative intensity changes across the \ha\  region 
as a function of wavelength (pixel by pixel).

\subsection{Reductions}
\label{sec:pix-reduc}

We could apply the time series analysis, as described in 
Section~\ref{sec:time-ser}, to the intensity of each pixel in our spectrum. 
However, this is computationally intensive and, since we are only interested 
in the amplitude of the principal mode, is unnecessary. 
Instead, the spectra were phase-binned at the principal frequency. 
The phase-binning was done so that Fast Fourier Transforms (FFTs) 
could be applied to the intensity variations, 
and therefore save calculation time 
for the measurement of the amplitude of the principal mode.

A sample of 4718 spectra from the Stromlo data set were used to produce 
the phase-binned spectra. 
Spectra which produced bad data points from the EW measurements 
(Section~\ref{sec:ew-reduc}) were not included. 
Each spectrum was linearly re-binned by a factor of 10 and shifted 
so that the centre of the \ha\ line was in the same wavelength 
position for each spectrum (determined by cross-correlation). 
This was similar to the reduction in Section~\ref{sec:ew-reduc}. 
Additionally, the intensities in each spectrum were slightly adjusted, 
in order to remove low-frequency variations 
of the mean intensity across the \ha\  region (filter~9, 6522\AA --6604\AA). 
This was necessary in order to avoid introducing noise at the principal 
frequency due to the low-frequency intensity variations. 
Using this adjustment, we obtained consistent results between 
the filter measurements on the individual spectra 
and on the phase-binned spectra.
The spectra were phase-binned at the 
principal frequency using a weighted mean in each phase-bin 
(25 phase-bins were used but this number was not critical).
Finally, a FFT was applied to the phase-binned series of intensities, to 
measure the amplitude and phase of the principal mode and its harmonics 
for each pixel.

\begin{figure*}
\epsfxsize=15.0cm
\centerline{\epsfbox{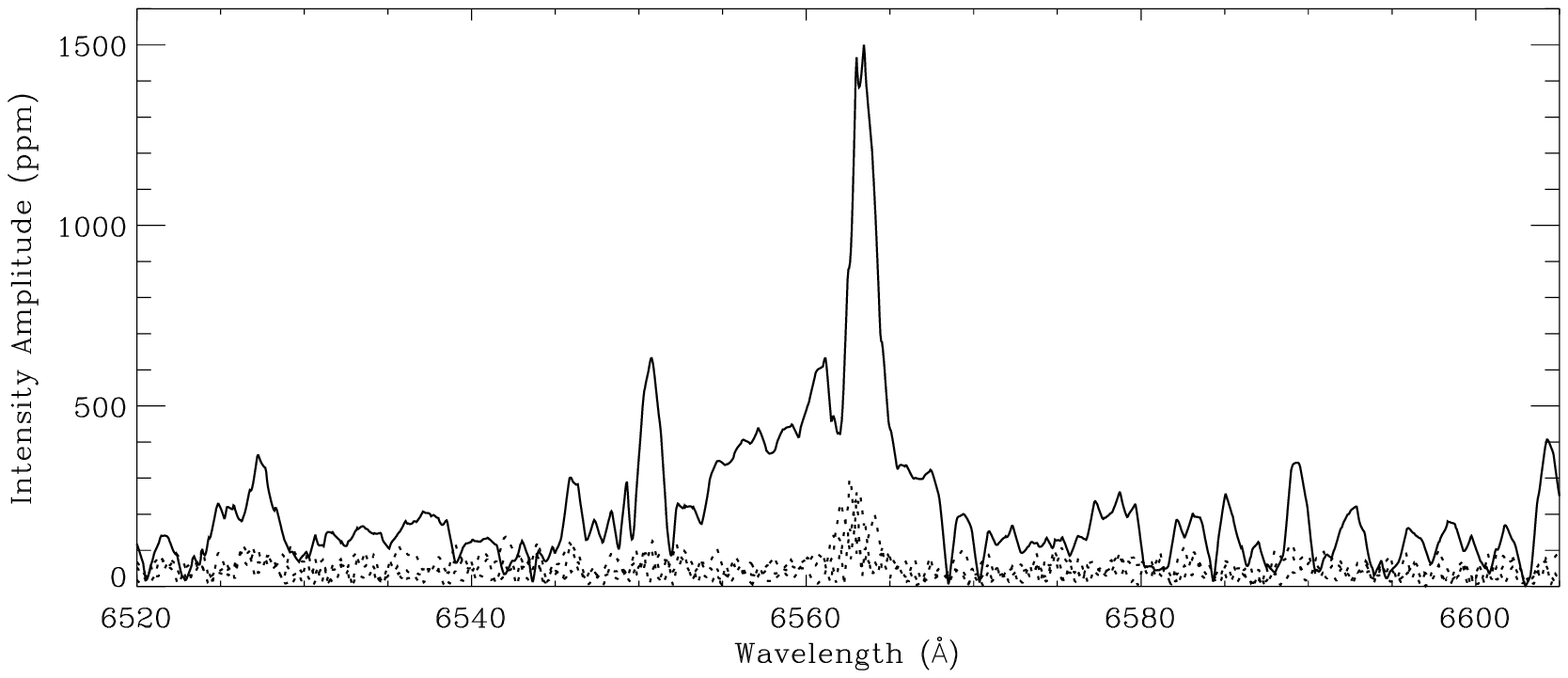}}
\caption{The solid line represents the relative intensity amplitude of the 
principal mode, while the other lines show three higher frequencies 
(harmonics).} 
\label{fig:amp-wave}
\epsfxsize=15.0cm
\centerline{\epsfbox{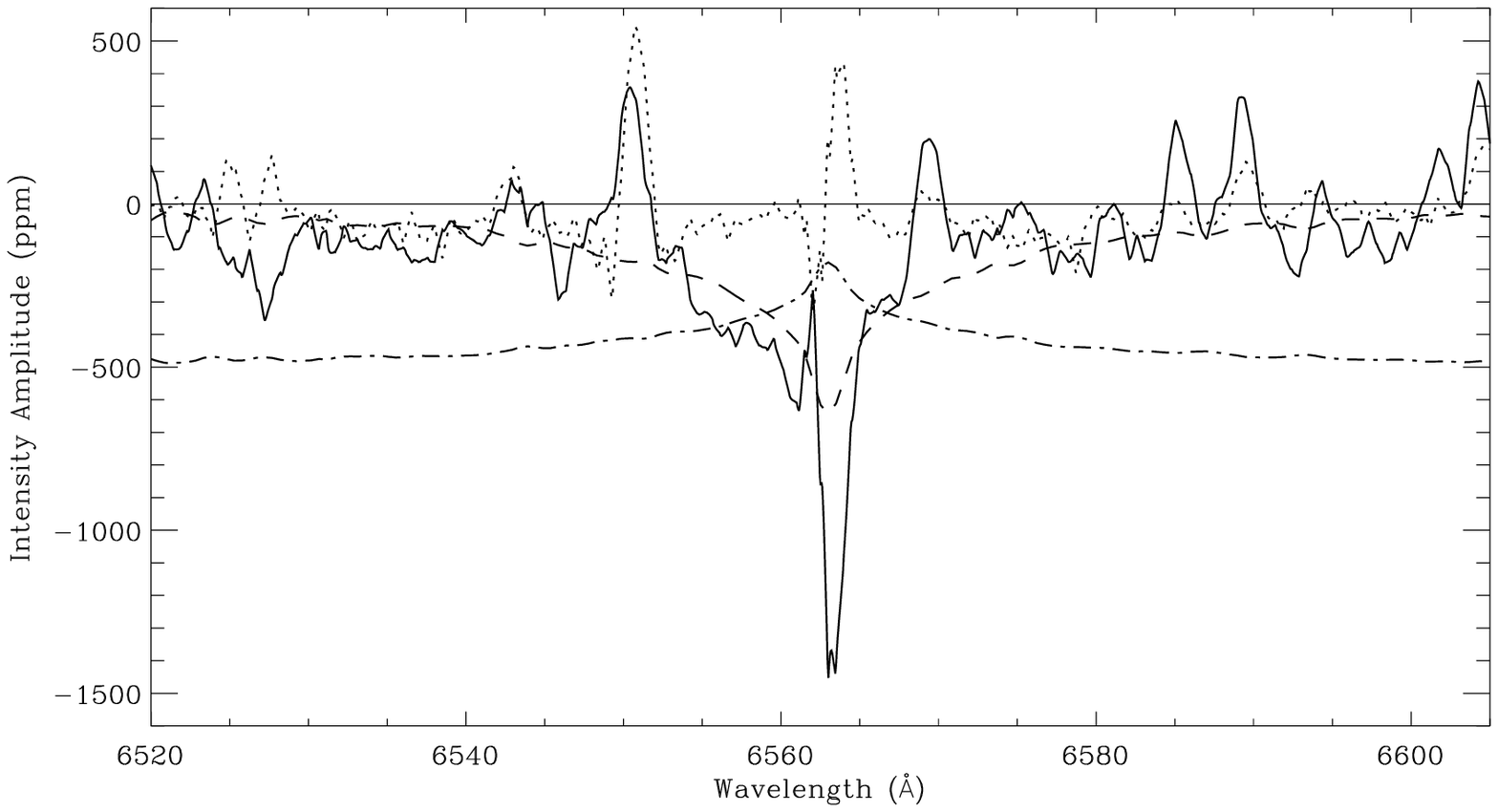}}
\caption{The solid line represents the relative intensity cosine amplitude 
of the principal mode, this component represents intensity changes in phase 
or in anti-phase with the photometric pulsation. 
The dotted line represents the sine amplitude. 
The dashed line is a theoretical amplitude with 
$\delta I = - 10^{-3} (C - I)$, 
and the dash-and-dotted line shows $\delta I = - 5 \times 10^{-4} I / C$ 
which represents approximately where the absolute flux amplitude 
would be zero (see text for details).} 
\label{fig:cos-sin-wave}
\end{figure*}

\subsection{Results}

The intensity amplitudes of the principal mode and harmonics, 
as a function of wavelength, are shown in Fig.~\ref{fig:amp-wave}. 
Note that a signal is expected in the first harmonic with an 
amplitude of about 8 percent of the principal mode (KSMT), 
with higher harmonics containing negligible signal. 
Therefore, the harmonics give a good indication of the noise level 
(the rms-noise is $\sim 60$\,ppm). 
{}From Fig.~\ref{fig:amp-wave}, 
it is clear that the simple description of the 
profile variation given in Section~\ref{sec:ew-res} is inadequate and 
that the metal lines have an impact on the intensity variations. 
For instance, 
the peak blueward of \ha\ could be caused by the Sr~I 6550.2\AA\ line, 
while the Fe~I 6569.2\AA\ line may also have an effect 
(this is more evident in the next figure). 
The signal from the core of \ha\ is strong while the signal from the 
wings is much weaker. 

To show the phase information, we have chosen to plot the cosine and sine 
components of the principal mode separately 
(see Fig.~\ref{fig:cos-sin-wave}). 
The phase of most pixels is near 0\degr\ or 180\degr\  and therefore most 
of the information is contained within the cosine component, 
which represents the relative intensity variations that are in 
phase or in anti-phase with the photometric pulsation. 
The amplitudes of this component are mostly negative which means that 
at maximum light, the relative intensity is at a minimum 
i.e.\ the EW of \ha\ is at a maximum. 

Two different simulations are also plotted: 
\begin{enumerate}
\item
The dashed line represents the relative intensity amplitude assuming 
$\delta I = - A (C - I)$, where $A$ is the EW amplitude of 1000\,ppm. 
This is the amplitude (cosine component) expected from the profile variation 
described in Section~\ref{sec:ew-res}. 
\item
The dash-and-dotted line represents $\delta I = - A I / C$ where 
$A$ is the continuum flux amplitude of 500\,ppm 
(Medupe \& Kurtz 1998). 
This line shows how 
the relative intensity amplitude (cosine component) 
would behave if the absolute flux 
in this region were constant (6520\AA\ -- 6605\AA), 
and the continuum flux outside this region were pulsating 
at the measured value of 500\,ppm. 
This is not a realistic simulation but we use it to estimate how the 
absolute flux in this region behaves. 
Where the solid line is above the dash-and-dotted line, 
the absolute flux is pulsating in phase with the continuum. 
We see that the absolute flux of the inner 10\AA\ of the \ha\ 
line is pulsating in anti-phase with the continuum. 
\end{enumerate}

\subsection{Discussion}
\label{sec:discuss-temp}

To evaluate these results, we have to consider systematic errors. 
Unlike for the bisector velocities, any systematic errors in the continuum 
level directly affect the EW and relative intensity measurements. 
The direction of any error would be the same at all wavelengths, 
equivalent to an offset in the cosine amplitude shown in 
Fig.~\ref{fig:cos-sin-wave}. 
{}From looking at pseudo-continuum regions outside \ha , 
we estimate the intensity amplitude of the continuum level 
to be 100\,ppm in the continuum fitted spectra 
(cosine component of $-$100\,ppm). 
This slightly changes the details of the \ha\  profile variation. 

Other systematic errors may be caused by 
residual Doppler shift signals that were not removed by the reduction process. 
It was not possible to remove such signals completely 
because of the variation in velocity amplitude and phase 
between different lines and between different heights in the \ha\  line.
This could account for some of the sine component of the amplitude. 

Ronan, Harvey \& Duvall (1991) measured the oscillatory signal in the Sun as a 
function of optical wavelength. 
One of their wavelength regions 
included the Balmer lines H$\gamma$ and H$\delta$. 
More recently, 
Keller et al.\ (1998) looked at the H$\beta$ line region in the Sun, 
using a similar technique. 
Both groups showed that 
the absolute flux oscillations in the Balmer lines 
were reduced to about 70 percent of the continuum signal. 
This differs sharply from \acir , where the \ha\  core absolute 
flux is pulsating in anti-phase with the continuum. 
This may be due to a pulsational temperature node in the atmosphere of \acir , 
as was indicated by the results of Medupe \& Kurtz (1998), 
and by the velocity node. 
The relationship between the temperature node and the velocity node is 
unclear with the present data. 

\section{Detection of Other Frequencies}

\begin{figure*}
\epsfxsize=15.0cm
\centerline{\epsfbox{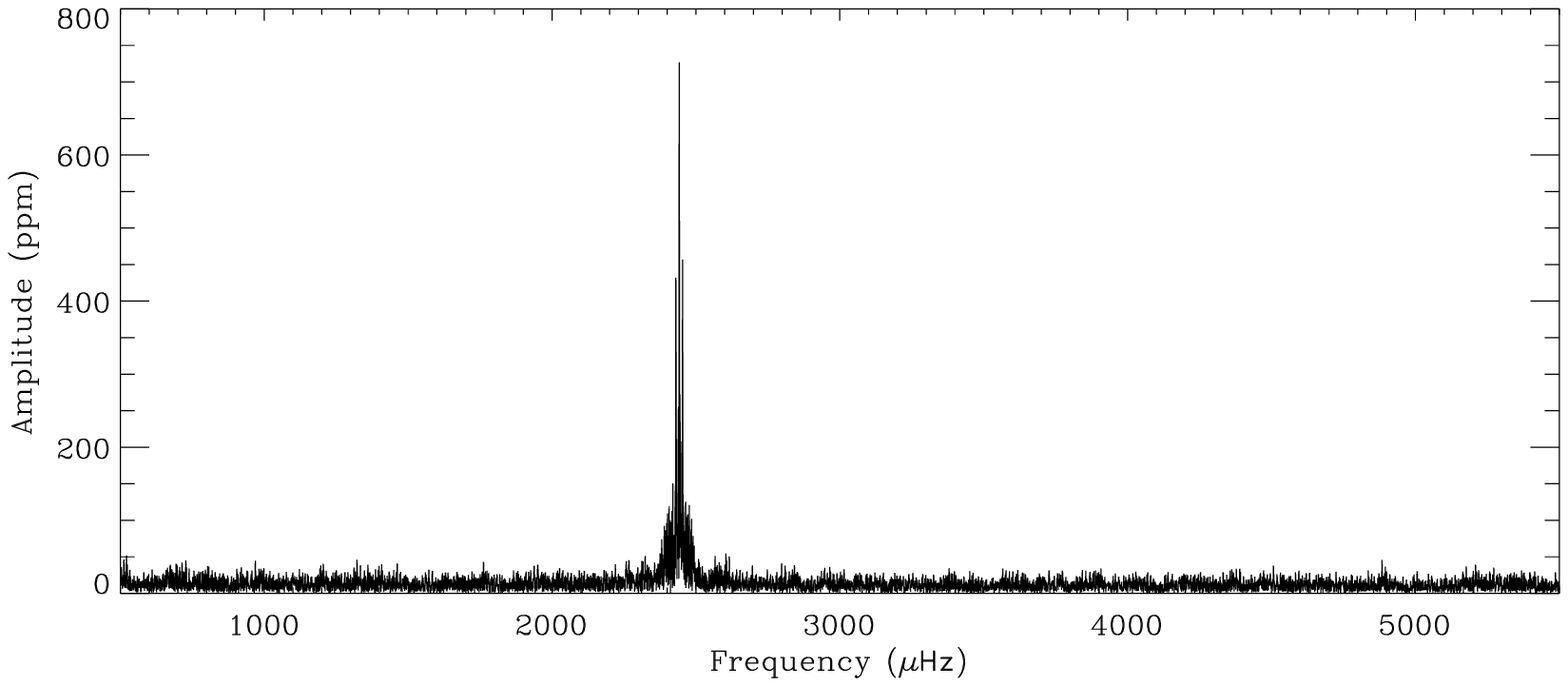}}
\caption{Amplitude ($\delta R_{cw}$) spectrum 
of the intensity ratio measurements.}
\label{fig:ew-ampl}
\end{figure*}

In order to detect weaker modes in \acir , we need an observable 
which has a high signal-to-noise ratio in the amplitude spectrum. 
Of the observables discussed so far, the one with 
the highest S/N for the principal mode is filter~2 from the EW measurements 
(S/N\,$= 35$, see Col.~8 of Table~\ref{tab:ew-meas}). 
However, for the measurements of intensity in different filters across \ha , 
the noise level is approximately the same for filters 2--10 
(20\,ppm, see Col.~5 of Table~\ref{tab:ew-meas}). 
This means that the noise is caused 
by variations in the continuum level and not by photon noise. 
Therefore, we can improve the S/N for the principal mode by dividing 
the intensity in one filter by another and thereby reducing errors 
caused by the continuum fit. 
The highest S/N ($= 51$) was obtained by dividing 
the intensity in filter~2 (FWHM $\sim 6$\AA) by filter~7 (FWHM $\sim 45$\AA).
We call this observable $R_{cw}$ (ratio of \ha\ core to wing intensity). 
This is analogous to narrow / wide H$\beta$ photometry. 

\begin{figure*}
\epsfxsize=15.0cm
\centerline{\epsfbox{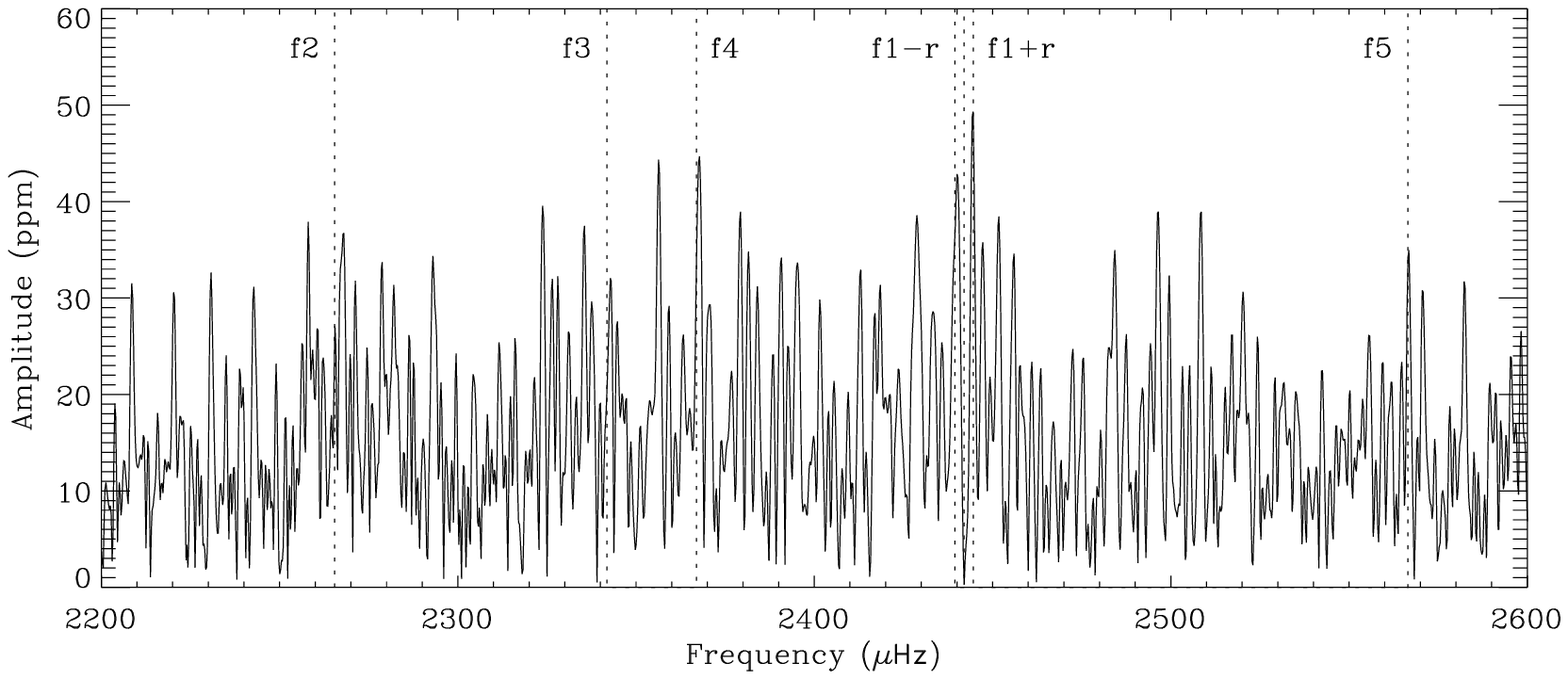}}
\caption{Amplitude spectrum after subtracting the principal frequency. 
The dotted lines represent the frequencies of modes detected by KSMT.} 
\label{fig:ew-pw-amp1}
\epsfxsize=15.0cm
\centerline{\epsfbox{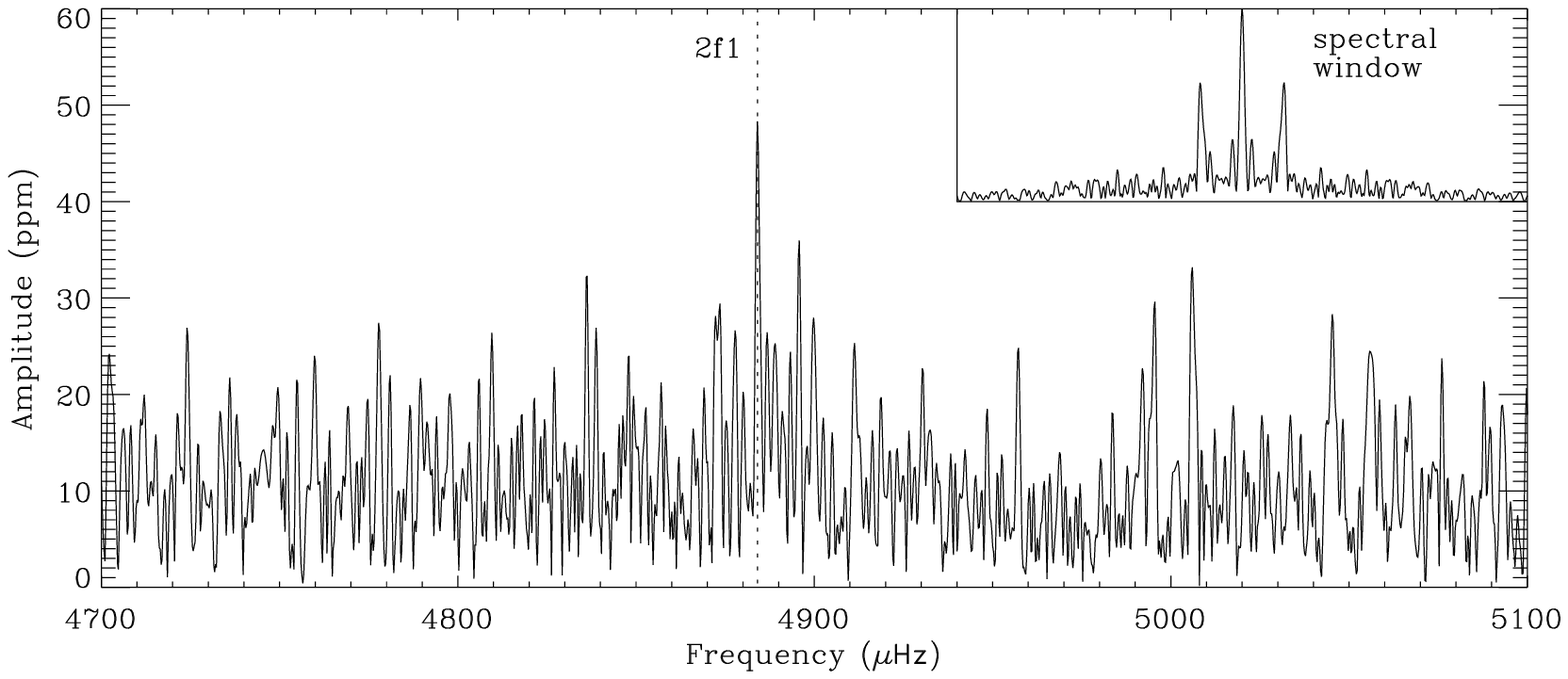}}
\caption{Same as Fig.~\ref{fig:ew-pw-amp1}, at a higher frequency. 
The dotted line represents the frequency of the first harmonic of the 
principal mode. 
The inset shows the spectral window on the same frequency scale.} 
\label{fig:ew-pw-amp2}
\end{figure*}

We produced a time series of these $R_{cw}$ measurements 
which was analysed in the same way as in Section~\ref{sec:time-ser}. 
The amplitude ($\delta R_{cw}$) of the principal mode 
was measured to be 727\,ppm, with an rms-noise level in the amplitude 
spectrum of 14\,ppm (see Fig.~\ref{fig:ew-ampl}). 
To search for other frequencies which were detected by KSMT 
(the same numbering is used), 
the principal pulsation mode was then subtracted from the time series 
to produce a pre-whitened amplitude spectrum 
(see Figs~\ref{fig:ew-pw-amp1}--\ref{fig:ew-pw-amp2}). 
In this spectrum, we have detected the modes $f_{4}$, $f_{5}$ and $2f_{1}$, 
and the rotational splitting of the principal mode ($f_{1} \pm r$), 
with amplitudes greater than $2.3 \times$rms-noise. 
We have not measured the frequencies but have used the values given by KSMT, 
except we have increased the frequencies of $f_{1}$ to $f_{5}$ by 0.03\mh\ 
for consistency with the change in the principal frequency, 
as measured during our observations (see Paper~I). 
The rotational splitting $r$ is taken as 2.59\mh , as measured by KSMT. 

\begin{table*}
\caption{The amplitudes ($\delta R_{cw}$) and phases of different modes 
measured using the intensity ratio of filter~2 and filter~7 
(see Table~\ref{tab:ew-meas}). 
Except for the principal mode ($f_{1}$), 
the measurements are made after pre-whitening with the principal frequency.} 
\label{tab:modes}
\begin{tabular}{lcrrrrcc} \hline
mode    &   freq.\rlap{$^a$}  & ampl. &  S/N\rlap{$^b$}  &
phase\rlap{$^c$}   & phase             & 
\multicolumn{2}{c}{relative amplitude}    \\
        & ($\mu$Hz)           & (ppm) &                  &
(\degr)            & error\rlap{$^d$}  & 
$\delta R_{cw}$\rlap{$^e$} & photometric\rlap{$^f$}  \\
          &          &      &       &       &     &       &        \\
$f_{2}$   &  2265.46 &   26 &   1.8 &  279  &  33 & 0.036 &  0.055 \\
$f_{3}$   &  2341.82 &   20 &   1.4 &  104  &  45 & 0.028 &  0.063 \\
$f_{4}$   &  2366.97 &   35 &   2.4 &  272  &  24 & 0.048 &  0.057 \\
$f_{1}-r$ &  2439.44 &   37 &   2.6 &   92  &  23 & 0.050 &  0.095 \\
$f_{1}$   &  2442.03 &  727 &  50.8 &  174  &   1 & 1.000 &  1.000 \\
$f_{1}+r$ &  2444.62 &   48 &   3.4 &  306  &  17 & 0.066 &  0.115 \\
$f_{5}$   &  2566.52 &   34 &   2.4 &   16  &  25 & 0.047 &  0.046 \\
$2f_{1}$  &  4884.06 &   48 &   3.4 &   76  &  17 & 0.066 &  0.077 \\ 
\hline
\end{tabular}
\begin{flushleft}
$^{a}$Frequencies taken from KSMT. See text for details.
\newline
$^{b}$rms-noise estimated to be 14\,ppm from the pre-whitened amplitude 
spectrum in the region 1100--4400\mh .
\newline
$^{c}$Phase measured 
with respect to a reference-point ($t_{0}$) at JD 2450215.07527, 
with the convention that a phase of 0\degr\ represents 
maximum of the observed variable. 
\newline
$^{d}$Error in the phase is taken to be arcsin\,(rms-noise/amplitude). 
\newline
$^{e}$Error of approximately 0.020 (rms-noise/amplitude of principal mode). 
\newline
$^{f}$Str\"omgren $v$ photometry (KSMT), with an error of approximately 0.007.
\newline
\end{flushleft}
\end{table*}

The amplitudes and phases of each mode are shown in Table~\ref{tab:modes}, 
with the last two columns showing 
the amplitude relative to that of the principal mode, 
both for our spectral data and for KSMT's photometric data. 
A possible significant difference is for the rotational side-lobes. 
Both side-lobes have a lower relative amplitude by more than 0.04 
in our data. 
This rotational splitting is caused by a variation in the amplitude 
of the principal mode during the rotation cycle ($P_{\rm rot} = 4.479$ days). 
Our results therefore suggest that the amplitude $\delta R_{cw}$ of 
the principal mode varies by less during the rotational cycle than 
the photometric amplitude.

\begin{figure}
\epsfxsize=9.5cm
\centerline{\epsfbox{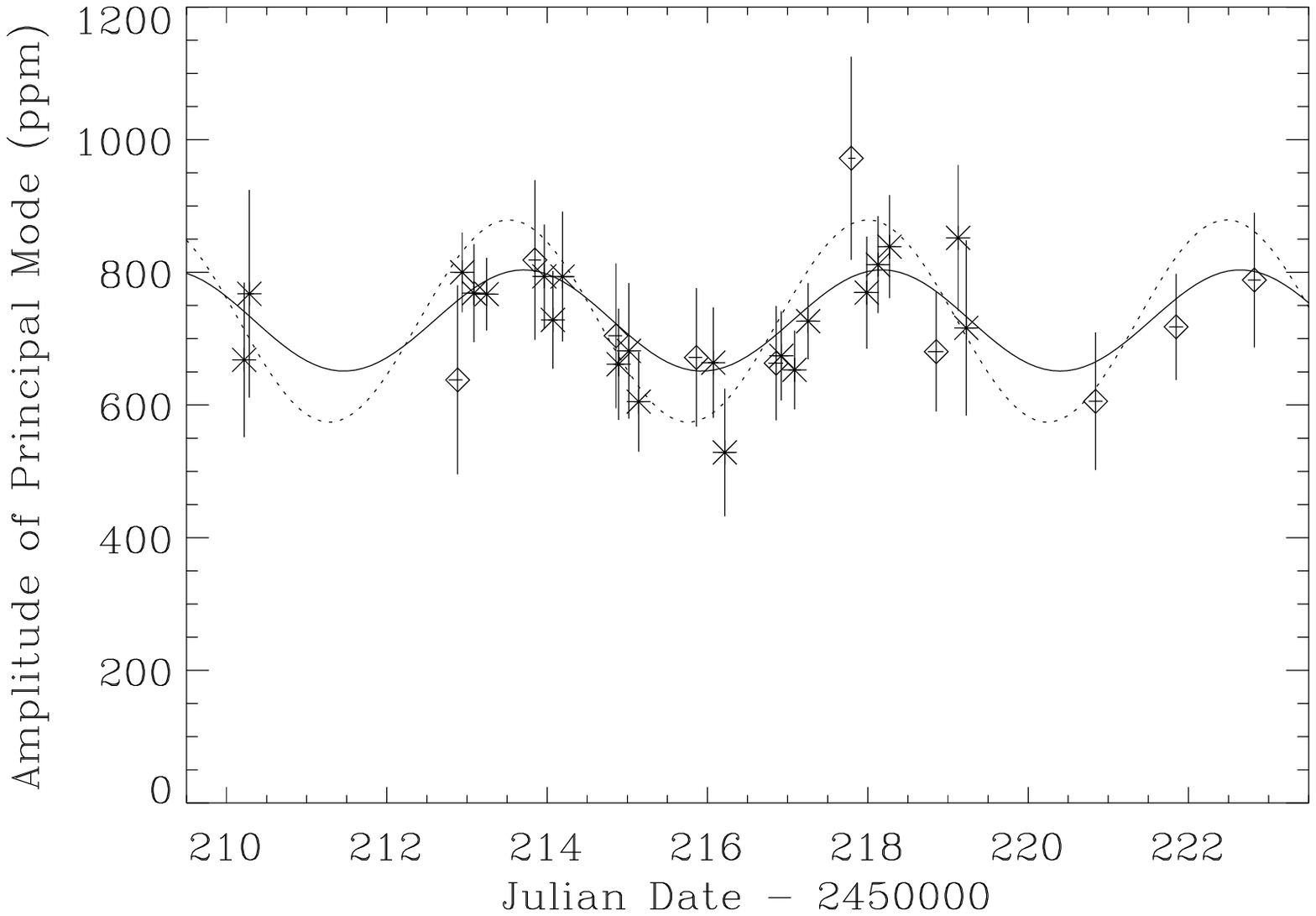}}
\caption{Amplitude ($\delta R_{cw}$) of the principal pulsation mode 
during separate time periods of approx.\ 4 hours. 
The asterisks represent the Stromlo data and 
the diamonds represent the La Silla data.
The error bar for each amplitude measurement is the rms-noise level.
See text for an explanation of the solid and dotted lines.}
\label{fig:ampl-jd}
\epsfxsize=9.5cm
\centerline{\epsfbox{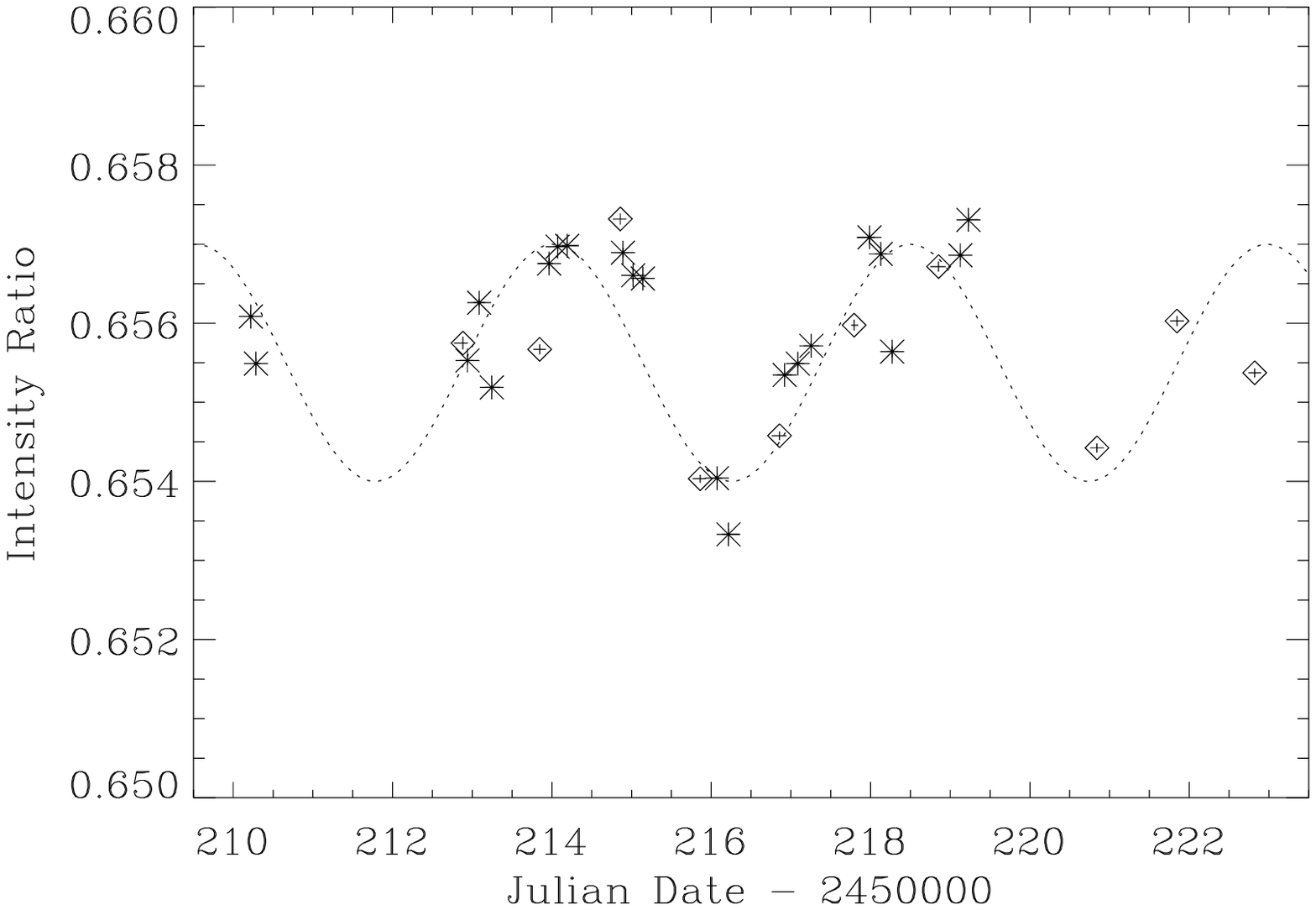}}
\caption{Average intensity ratio ($R_{cw}$) 
during separate time periods of approx.\ 4 hours. 
The asterisks represent the Stromlo data. 
The diamonds represent the La Silla data, which have been increased by 
0.4 percent to match the Stromlo data. 
The dotted line shows a sine-wave with a period equal to the rotation 
period of the star.}
\label{fig:ratio-jd}
\end{figure}

To examine this, we divided the time series into 32 shorter time periods of 
about 4 hours each. 
In Fig.~\ref{fig:ampl-jd}, we show the amplitude $\delta R_{cw}$ of the 
principal mode for each time period. 
The solid line shows the amplitude variation expected from the measurements 
of $f_{1}$, $f_{1}-r$ and $f_{1}+r$ in the amplitude spectrum of the 
complete time series (Table~\ref{tab:modes}). 
The dotted line shows the expected amplitude variation based on the 
full range variation of 21 percent and the ephemeris from the photometric 
results (KSMT).
There is good agreement between the amplitude maximum of our fit and 
the ephemeris of KSMT, 
which supports the accuracy of the rotation period derived by them. 
Both fits shown in Fig.~\ref{fig:ampl-jd} appear to be consistent 
with the results of the analysis using the shorter time periods. 
Therefore, we cannot say with certainty that 
the percentage amplitude variation is less in our data 
than the variation measured by KSMT. 
Such a difference, if it exists, may be due to limb-darkening.

The average value of $R_{cw}$ itself in each time period varied slightly 
during the observation period and was slightly different between 
the Stromlo and La Silla data sets (see Fig.~\ref{fig:ratio-jd}). 
The difference in the value between the two data sets was about 0.4 percent 
and was due to the inexact matching of the filters. 
The variation with time seen in Fig.~\ref{fig:ratio-jd} is probably caused 
by the rotation of the star, 
which is slightly inhomogeneous in surface brightness. 
This variation does not significantly affect the measurement of 
$\delta R_{cw}$ at pulsation frequencies. 

Variation of Balmer line profiles with rotation is common in Ap stars. 
For example, 
Musielok \& Madej (1988) investigated 22 Ap stars of which 17 showed 
variation of the H$\beta$ index with a typical amplitude of 0.02\,mag 
(19000\,ppm). 
Our observed $\delta R_{cw}$ amplitude of about 1500\,ppm
at the rotation frequency in \acir\  corresponds to a much smaller 
Balmer line variation than is typical of Ap stars, even if we allow for 
some difference between the H$\beta$ index and $R_{cw}$.
To test this, we calculated an intensity ratio using filters across \ha\ 
of similar widths to those used for the H$\beta$ index 
(FWHMs $\sim 40$\AA\ and $\sim 150$\AA). 
The rotational amplitude of this intensity ratio was 1000--1500\,ppm, 
which is more than a factor of ten smaller than 
the 0.02\,mag found by Musielok \& Madej in other Ap stars. 

\section{Conclusions}

The existence of a radial standing wave node 
of the principal pulsation mode has been put on a firm 
footing by the velocity phase reversal in the \ha\ line. 
However, there is only a 140\degr\ velocity phase shift 
between the \ha\ core and higher in the line. If we were seeing a pure 
standing wave in the atmosphere then we would expect a 180\degr\ phase shift. 
A travelling wave component, blending or the systematic effects 
described in Section~\ref{sec:simul} may cause the discrepancy.
Additionally, 
many of the metal lines studied in Paper~I show phases that are neither 
in phase or in anti-phase with the majority of the lines. 
We proposed that some of the phases were anomalous and did not represent 
velocity phases due to blending effects. 
Therefore, in a time series of less blended (higher resolution) spectra, 
there should be a clearer distinction between those lines that are formed 
above and below the velocity node and fewer lines with anomalous phases. 

Analysis of the \ha\ profile during the principal pulsation cycle, 
shows a large change in relative intensity of the core of the line. 
This means that the absolute flux of the core is pulsating in 
anti-phase with the continuum, 
indicating a pulsational temperature node in the atmosphere, 
as suggested by Medupe \& Kurtz (1998) 
on the basis of multi-colour photometry. 
However, Medupe, Christensen-Dalsgaard \& Kurtz (1998) show that 
non-adiabatic effects can explain the photometric results without 
the need for a node. 
Perhaps, non-adiabatic effects have a significant effect on the 
\ha\ profile changes in \acir . 

The next stage in the analysis of the mode dynamics in the atmosphere 
could be the use of high resolution (R$\sim$50000) spectroscopy to study 
the velocities, bisectors and EW changes of unblended metal lines, 
combined with model atmospheres to calculate the 
formation height of the lines. 
This may allow us to decide whether there is travelling wave component, 
to relate the velocity changes to the temperature changes, 
to measure any surface or vertical inhomogeneities 
in the distribution of certain elements, 
and to test non-adiabatic non-radial pulsation equations. 

By defining an observable which detects the relative intensity changes 
in the core of the \ha\ line,  $R_{cw}$, we detected some of the weaker 
modes in \acir . 
This is probably the first detection, in roAp stars, 
of modes with photometric ($B$) amplitudes of less than 0.2\,mmag, 
using a spectroscopic technique. 
The ratio between spectroscopic amplitudes and photometric amplitudes can be 
used to identify the $\ell$ value of different oscillations modes, 
as has been done recently by Viskum et al.\ (1998) for 
the $\delta$~Scuti star FG~Vir. 
We do not have enough S/N to identify the modes in \acir , 
but we show the potential for using this mode identification technique.

\acir\ offers one of the best chances for theoreticians to model a star 
where magnetic fields are important both for its evolution and pulsation, 
because of the detailed information that can be gained from its oscillation 
modes using spectroscopy and photometry. 
Additionally, 
there is already a well determined luminosity using Hipparcos and, 
when angular diameter measurements are made, there will also be 
direct measurements of the effective temperature and radius of the star.

\section{Acknowledgements}

Many thanks to Jaymie Matthews and Don Kurtz, 
with whom we had fruitful discussions, 
and to Friedrich Kupka and DK for providing data on \acir . 
Thanks also to Thebe Medupe and Lawrence Cram for help with 
the model atmospheres. 
We are grateful to Michael Bessell for his help and advice in setting 
up the instrument at Mt.~Stromlo, 
and to the Director of Mt.~Stromlo Observatory and 
the Danish Time Allocation Committee for the telescope time. 
We acknowledge the Harvard ADS Abstract Service 
as an invaluable aid in finding relevant papers. 

This work was carried out while IKB was in receipt of an 
Australian Postgraduate Award, 
and was also supported by funds from the Australian Research Council, 
the Danish Natural Science Research Council through its Center for 
Ground-based Observational Astronomy, 
and the Danish National Research Foundation through its establishment 
of the Theoretical Astrophysics Center.

\label{lastpage}


\begin{thebibliography}{99}

\bibitem{b00} Baldry I.K., Bedding T.R., Viskum M., Kjeldsen H., Frandsen S., 
   1998a, MNRAS, 295, 33 (Paper~I)
\bibitem{b01} Baldry I.K., Bedding T.R., Viskum M., Kjeldsen H., Frandsen S., 
   1998b, in Deubner F.-L., Christensen-Dalsgaard J., Kurtz D.W., eds, 
   Proc.\ IAU Symp.\ 185, 
   New Eyes to See Inside the Sun and Stars. 
   Kluwer, Dordrecht, p.~309
\bibitem{b02} Bedding T.R., Kjeldsen H., Reetz J., Barbuy B., 1996, 
   MNRAS, 280, 1155 
\bibitem{b25} Brown T.M., Gilliland R.L., 1994, ARA\&A, 32, 37
\bibitem{b22} Christensen-Dalsgaard~J., 1998, 
   Lecture Notes on Stellar Oscillations. Univ.\ Aarhus \\
   ({\tt http://www.obs.aau.dk/\~{}jcd/oscilnotes/})
\bibitem{b03} Deubner F.-L., Waldschik Th., Steffens S., 1996, 
   A\&A, 307, 936 
\bibitem{ESA} ESA, 1997, The Hipparcos and Tycho Catalogues. ESA SP-1200
\bibitem{GSH} Gautschy A., Saio H., Harzenmoser H., 1998, MNRAS, submitted
\bibitem{b14} Hatzes A.P., 1996, PASP, 108, 839 
\bibitem{b17} Keller C.U., Harvey J.W., Barden S.C., Giampapa M.S., Hill F., 
   Pilachowski C.A., 1998, in Deubner F.-L.\ et al., eds, 
   New Eyes to See Inside the Sun and Stars. Kluwer, Dordrecht, p.~375
\bibitem{b04} Kjeldsen H., Bedding T.R., 1995, A\&A, 293, 87 
\bibitem{b11} Kjeldsen H., Bedding T.R., Viskum M., Frandsen S., 1995, 
   AJ, 109, 1313 
\bibitem{b10} Kupka F., Ryabchikova T.A., Weiss W.W., Kuschnig R., Rogl J.,
   Mathys G., 1996, A\&A, 308, 886 
\bibitem{b24} Kurtz D.W., 1982, MNRAS, 200, 807
\bibitem{b05} Kurtz D.W., Sullivan D.J., Martinez P., Tripe~P., 1994, 
   MNRAS, 270, 674 (KSMT) 
\bibitem{b06} Kurucz R.L., 1979, ApJS, 40, 1 
\bibitem{b07} Medupe R., Kurtz D.W., 1998, MNRAS, 299, 371
\bibitem{b27} Medupe R., Christensen-Dalsgaard J., Kurtz D.W., 1998, 
   in Bradley P.A., Guzik J.A., eds, 
   ASP Conf.\ Ser.\ Vol.\ 135, 
   A Half Century of Stellar Pulsation Interpretations.
   Astron.\ Soc.\ Pacific, San Francisco, p.~197
\bibitem{b23} Musielok B., Madej J., 1988, A\&A, 202, 143 
\bibitem{b16} Ronan R.S., Harvey J.W., Duvall T.L. Jr., 1991, 
   ApJ, 369, 549 
\bibitem{b19} Schmidt-Kaler Th., 1982, in Schaifers K., Voigt H.H., eds, 
   Landolt-B\"ornstein: Numerical Data and Functional Relationships in 
   Science and Technology, Group~VI, Vol.~2b. 
   Springer-Verlag, Berlin, p.~451
\bibitem{b21} Shibahashi H., Takata M., 1993, PASJ, 45, 617 
\bibitem{b26} Viskum M., Kjeldsen H., Bedding T.R., Dall T.H., Baldry I.K., 
   Bruntt H., Frandsen S., 1998, A\&A, 335, 549

\end{thebibliography}
\end{document}